# Fokker-Planck equation for the particle size distribution function in KJMA transformations.

By

Massimo Tomellini

*Dipartimento di Scienze e Tecnologie Chimiche Università di Roma Tor Vergata Via della Ricerca Scientifica 00133 Roma, Italy*

**Abstract**

The Fokker-Planck (FP) equation has been derived for describing the temporal evolution of the particle size probability density function (PDF) for KJMA (Kolmogorov-Johnson-Mehl-Avrami) transformations. The classical case of transformations with constant rates of both nucleation and growth, in 3D space, has been considered. Integration of the equation shows that the PDF is given by the superposition of one-parameter Gamma distributions with time dependent mean size given by the KJMA theory. The asymptotic behavior of the FP solution offers a demonstration of the conjecture, previously proposed by Pineda et al [E. Pineda, P. Bruna, D. Crespo, Phys. Rev. E **70** (2004) 066119], according to which the set of nuclei formed at the same time are Gamma-distributed, with parameter depending on nucleus birth time. Computer simulations of the transformation with constant nucleation and growth rates, do show that the temporal evolution of the PDF, in the volume domain, is in good agreement with the Johnson-Mehl PDF. The approach based on the FP equations provides a particle size PDF that exhibits such behavior.

**1-Introduction**

Phase transformations by random nucleation and irreversible isotropic growth are well described by the Kolmogorov-Johnson-Mehl-Avrami model (KJMA) [1-5]. The approach correctly describes the impingement growth-mode where, because of collisions among particles, the growth of nuclei stops at the common interface. This mean-field theory provides the temporal evolution of the mean volume of particles, $\bar{x}(t,\tau)$, as a function of actual time, $t$, and birth time, $\tau$, and the volume fraction of the transformed phase. However, the classical KJMA model, being a mean-field theory, ignores fluctuations around the mean particle size. In this approach, the particle size probability density function (PDF) is the image of the nucleation rate at the nuclei time of birth. The functional dependence of the PDF with the particle size is eventually obtained by inverting $\bar{x} = \bar{x}(t,\tau)$. This size distribution function is usually called Johnson and Mehl (JM) PDF [3].

To model the temporal evolution of the PDF in nucleation and growth transformations, kinetic models based on rate equations, and Fokker-Planck-type equations, have been employed [6-13]. The importance of fluctuations, in KJMA compliant transformations is evident in the simultaneous nucleation process where the final PDF (at time $t \to \infty$), in the volume domain, is the Gamma distribution function typical of the Poisson Voronoi (PV) tessellation. This distribution deviates, significantly, from a Dirac delta [14-16]. In fact, by neglecting fluctuations around the mean particle size, the shape of the PDF would mimic the nucleation rate profile, in this case a Dirac delta. Fluctuations in particle size are due to the stochastic nature of the collision events among nuclei. Concerning the temporal evolution of the PDF for simultaneous nucleation, an analytical approach has been presented in [9]. The method is based on the computation of the probability, $P_k$, a particle collides with $k$ neighboring particles in each spatial configuration. The model does not give the asymptotic PDF (at $t \to \infty$) owing to the increased number of configurations needed to estimate the PDF. Application of the FP equation for modeling the evolution of the PDF in site-saturated nucleation has been discussed in [17].

In addition to the simultaneous nucleation, the classical KJMA approach can also be used to model situations with constant rates of both nucleation and growth. For this transformation the final PDF is that of the Johnson and Mehl (JM) tessellation. Applications of the JM cell-size distribution function have been reported in refs.[18,19]. These approaches are based on the conjecture, proposed by Pineda, Bruna and Crespo [18], according to which the PDF is given by the superposition of one-parameter Gamma PDFs with parameters estimated by means of the statistical method developed by Gilbert [20]. The validity of this conjecture was also verified via computer simulations [18].

Although not compliant with the classical KJMA transformation, it is worth quoting the work by Hömberg et al on the evolution of the particle size PDF in steel [12]. The authors solved the Fokker Planck (FP) equation for diffusional-type growth to obtain the lognormal distribution in agreement with experimental reports.

It is worth recalling that in the FP equation, fluctuations are due to the second order derivative term in the spatial variable. In a previous paper we studied the role of fluctuations in KJMA compliant transformations in 3D space [17]. It was shown that, in the volume domain, fluctuations can be neglected, and the PDF reduces to the Johnson and Mehl PDF (JM-PDF) that is given in terms of the nucleation rate and time dependence of the mean particle size [3]. The present work aims to solve the FP equation for the particle size PDF, using the growth rate derived from the KJMA approach. In addition, the solution of the FP equation provides a rationale for the negligible role of fluctuations in the FP equation leading to the temporal evolution of the JM-PDF. As we will show, this can be traced back to the conjecture proposed by Pineda et al [18] according to which nucleation events between time $\tau$ and $\tau + d\tau$ give rise, at the final time of the transformation, to a set of particles distributed in size according to a Gamma PDF. One of the goals of the present paper is to provide a demonstration of this conjecture by solving the FP equation for classical KJMA transformation.

The article is divided as follows. Section 2.1 is devoted to summarizing the mean-values of the kinetic quantities accessible by the KJMA theory. The FP equation for the PDF of KJMA compliant transformations is derived in section 2.2. In the same section the FP equation is integrated, analytically, providing a demonstration of the conjecture by Pineda, Bruna and Crespo. Additionally, application of the approach to the non-isothermal transformations undergoing a constant rate of heating, is also discussed.

**2-Results and discussion**

*2.1. Classical KJMA theory*

The KJMA model gives the kinetics of phase transformation by nucleation and growth for random distribution of nuclei and impingement among growing nuclei. The fraction of volume transformed at time $t'$, $X(t')$, is given by [1-5]

$$X(t') = 1 - e^{-X_e(t')}, \qquad (1a)$$

where $X_e(t') = \int_0^{t'} I(t'') x_e(t',t'') dt''$ is the extended volume. In this expression, $I(t'')$ is the nucleation rate (phantom included) and $x_e = g_D r^D$ is the extended volume of the nucleus with $D$ the space dimension, $g_D$ a shape factor (equal to $2, \pi,$ and $\frac{4\pi}{3}$ for $D=1, 2$ and 3, respectively) and $r \equiv r(t',t'')$ the nucleus growth law. In the following, we confine our study to the case of constant nucleation rate and linear growth, where $I$ and $G = dr/dt'$, are both constants. Using the reduced variables, $t = \omega t'$, with $\omega = \left(\frac{I g_D G^D}{D+1}\right)^{1/(D+1)}$, eqn.1a becomes

$$X(t) = 1 - e^{-t^{D+1}}. \qquad (1b)$$

Additionally, the KJMA approach provides the time dependence of the mean particle size, $\bar{x}$, according to the differential equation [1]

$$\frac{d\bar{x}(t,\tau)}{dt} = \frac{(1-X(t))}{(1-X(\tau))} \frac{dx_e(t,\tau)}{dt}, \qquad (2a)$$

where $\tau$ ($\tau < t$) is the birth time of the nucleus. By defining the reduced volume $\bar{v} = \frac{\bar{x}}{\lambda^D}$, with $\lambda = (D g_D)^{\frac{1}{D}} \frac{G}{\omega}$, eqns.1, 2 yield

$$\frac{d\bar{v}(t,\tau)}{dt} = \frac{e^{-t^{D+1}}}{e^{-\tau^{D+1}}} (t-\tau)^{D-1}, \qquad (2b)$$

with solution

$$\bar{v}(t,\tau) = e^{\tau^{D+1}} \int_\tau^t e^{-z^{D+1}} (z-\tau)^{D-1} dz. \qquad (3a)$$

The first modeling of the PDF for KJMA transformations was developed by Johnson and Mehl for constant rates of nucleation and growth [3]. The JM-PDF is derived by neglecting fluctuations around the mean size of each population of $\tau$-nuclei. This PDF is attained from eqn.3a by assuming the spread of the distribution due to the nucleation process only. The JM-PDF, $f_{JM}$, is computed knowing the kinetics of the mean size of nuclei according to [17]

$$f_{JM}(v,t) = N^{-1} \frac{e^{-\tau(v,t)^{D+1}}}{\left|\frac{\partial \bar{v}(t,\tau)}{\partial \tau}\right|_{\tau(v,t)}} \qquad (3b)$$

with $\tau(v,t)$ root of the equation $v = \bar{v}(t,\tau)$ and $N = \left(\int_0^t e^{-\xi^{D+1}} d\xi\right)$ is a normalization factor (see appendix for details).

For transformations in 3D space, eqn.3a becomes

$$\bar{v}(t,\tau) = \frac{1}{4} e^{\tau^4} \left(\tau^2 \Gamma\left(\frac{1}{4}, \tau^4, t^4\right) - 2\tau \Gamma\left(\frac{2}{4}, \tau^4, t^4\right) + \Gamma\left(\frac{3}{4}, \tau^4, t^4\right)\right), \qquad (4)$$

where $\Gamma(a,b,c)$ is the generalized incomplete Gamma function[1]. The behavior of $\bar{v}(t,\tau)$ (eqn.4) has been reported in Fig.1 as a function of $t$ and for several values of $\tau$. These sigma-shaped curves are in excellent agreement, at given $\tau$, with stretched exponential functions of the form

$$\bar{v}(t,\tau) = \bar{v}_{\tau,\infty}\left(1 - e^{-b(\tau)[t-\tau]^{\kappa(\tau)}}\right) \qquad (5)$$

with $b$ and $\kappa$, $\tau$-dependent parameters and $\bar{v}_{\tau,\infty} = \bar{v}(\infty,\tau)$. It follows that, in terms of the new variable $\xi_\tau \equiv \xi(t;\tau) = b(\tau)[t-\tau]^{\kappa(\tau)}$, the temporal evolution of the mean volume of the "$\tau$-nuclei population" (eqn.5), $\bar{v}_\tau = \bar{v}_\tau(\xi_\tau) = \bar{v}_{\tau,\infty}(1 - e^{-\xi_\tau})$, satisfies the differential equation

$$\frac{d\bar{v}_\tau}{d\xi_\tau} = \bar{v}_{\tau,\infty} - \bar{v}_\tau, \qquad (6)$$

---

[1] The generalized incomplete Gamma function is defined as $\Gamma(a, z_0, z_1) = \int_{z_0}^{z_1} x^{a-1} e^{-x} dx$. Besides, $\Gamma(a) \equiv \Gamma(a, 0, \infty)$.

where the subscript emphasizes the birth time of the class of nuclei ($\tau$-nuclei). Eqn.6 is at the basis of the development of the Fokker-Planck equation for KJMA transformations discussed in the next section.

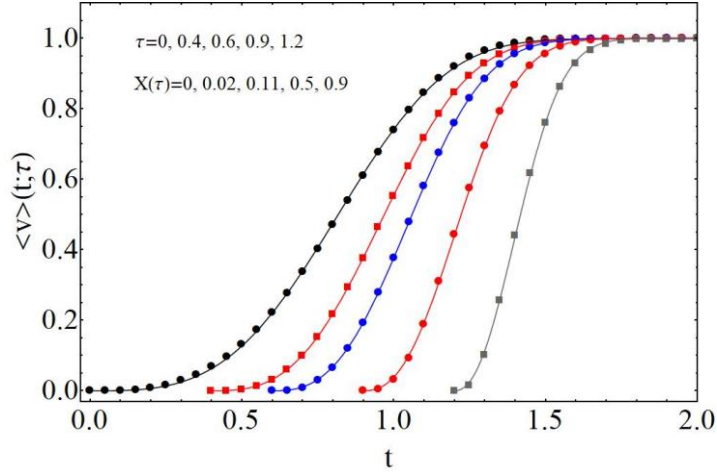

**Fig.1** Solid symbols: normalized mean nucleus volume ($<v> = \frac{\bar{v}(t,\tau)}{\bar{v}_{\tau,\infty}}$) as a function of actual time, $t$ and for several values of nucleus birth time, $\tau$. In the panel, the values of $\tau$ and $X(\tau)$, namely the transformed volume fraction at time $\tau < t$, are displayed. Solid lines are the best fit of eqn.5 to the kinetics of the mean volume (eqn.4).

*2.2. Fokker- Planck equation for KJMA transformation.*

The Fokker-Planck equation (also named Kolmogorov's second equation) is written as [17]

$$\frac{\partial F(\tau,y|t,x)}{\partial t} = -\frac{\partial [a(t,x)F(\tau,y|t,x)]}{\partial x} + \frac{1}{2}\frac{\partial^2 [b(t,x)F(\tau,y|t,x)]}{\partial x^2}, \qquad (7a)$$

where $F(\tau,y|t,x)dx$ is the probability that the value of the stochastic variable is between $x$ and $x + dx$ at time $t$, given that its value is $y$ at time $\tau$ ($\tau < t$). In eqn.7a, $a(t,x)$ is the mean value of the rate of variation of the stochastic variable at $(t,x)$, and $b(t,x)$ the mean value of the rate of variation of the square of the stochastic variable. For the system investigated here, the stochastic variable is the nucleus volume and $\tau$ is set equal to the birth time of nuclei when $x = 0$. The Johnson-Mehl PDF is attained from eqn.7a for $b=0$, that is, by neglecting fluctuations around the mean size of nuclei that start growing at time $\tau$. This PDF has been discussed in detail in ref.[19] and its functional form derived in the Appendix.

For progressive nucleation, let us consider a single class of $\tau$-nuclei where, it is worth recalling, $\tau$ enters the FP equation as a parameter. Using the $\xi_\tau$ and $v$ variables, eqn.7a becomes,

$$\frac{\partial f_\tau(\xi_\tau, v)}{\partial \xi_\tau} = -\frac{\partial [A_\tau(\xi_\tau, v) f_\tau(\xi_\tau, v)]}{\partial v} + \frac{1}{2}\frac{\partial^2 [B_\tau(\xi_\tau, v) f_\tau(\xi_\tau, v)]}{\partial v^2} \qquad (7b)$$

where $f_\tau(\xi_\tau, v) = F(\tau, 0|t(\xi_\tau, \tau), v)$, $A_\tau(\xi_\tau, v) = \frac{\partial t(\xi_\tau, \tau)}{\partial \xi_\tau} a[t(\xi_\tau, \tau), v]$ and $B_\tau(\xi_\tau, v) = \frac{\partial t(\xi_\tau, \tau)}{\partial \xi_\tau} b[t(\xi_\tau, \tau), v]$. From eqn.7b the following relationships hold for the mean value of the first and second moments of the $v$ variable [21] (see also the appendix)

$$\frac{d\bar{v}_\tau(\xi_\tau)}{d\xi_\tau} = \langle A_\tau(\xi_\tau, v) \rangle \qquad (8a)$$

and

$$\frac{d\Psi_\tau}{d\xi_\tau} = 2\langle (v - \bar{v}_\tau) A_\tau(\xi_\tau, v) \rangle + \langle B_\tau(\xi_\tau, v) \rangle, \qquad (8b)$$

where $\Psi_\tau = \langle v^2 - \bar{v}_\tau^2 \rangle$ and the $\tau$-dependent averages are taken over the $f_\tau$-PDF. From eqn.6 and eqn.8a it follows that, in the FP equation for the KJMA model, $A_\tau(\xi_\tau, v) = A_\tau(v) = \bar{v}_{\tau,\infty} - v$. Concerning the fluctuation term, $B_\tau(\xi_\tau, v)$, let us consider a linear behavior of $B_\tau$ of the form $B_\tau(\xi_\tau, v) = B_\tau(v) = \frac{2v}{\gamma_\tau}$, with constant $\gamma_\tau$. This assumption stems from the consideration that the larger the nucleus volume (at given $\tau$) the greater the chance to be subjected to impingement events[2]. Therefore, for the $\tau$-class of nuclei the FP equation reads

---

[2] The first order expansion of the diffusion term is of the form $B \approx v$. See also the discussion at the end of this section.

$$\frac{\partial f_\tau(\xi_\tau, v)}{\partial \xi_\tau} = -\frac{\partial\left[(\bar{v}_{\tau,\infty} - v)f_\tau(\xi_\tau, v)\right]}{\partial v} + \frac{1}{2\gamma_\tau}\frac{\partial^2\left[2vf_\tau(\xi_\tau, v)\right]}{\partial v^2}. \quad (9)$$

In the demonstration that follows, to simplify the notation the $\tau$-subscripts will be omitted in eqn.9. For each value of the $\tau$ parameter eqn.9 can be solved employing the method based on the orthogonal-polynomial expansion of the PDF [22]. According to this approach, for the initial condition $f(v_0|0, v) = \delta(v - v_0)$, the solution is given by the series

$$f(v_0|\xi, v) = q(v) \sum_{n=0}^{\infty} e^{-\lambda_n \xi} \varphi_n(v) \varphi_n(v_0), \quad (10)$$

where $q(v)$ is the steady state solution of the PDF (i.e. the FP solution in the limit $\frac{\partial f(\xi,v)}{\partial \xi} \to 0$) and the polynomials, $\varphi_n(v)$, satisfy the Sturm-Liouville equation with discrete eigenvalues, $\lambda_n$ [3]:

$$\frac{1}{2}\frac{d}{dv}\left(B(v)q(v)\frac{d\varphi_n(v)}{dv}\right) + \lambda_n q(v)\varphi_n(v) = 0 . \quad (11)$$

The function $q(v)$ satisfies the differential equation

$$\frac{dq(v)}{dv} = \frac{1}{B(v)}\left(2A(v) - \frac{dB(v)}{dv}\right)q(v) \quad (12a)$$

Which upon using the definition of A and B becomes

$$\frac{dq(v)}{dv} = \frac{\gamma}{2v}\left[2(\bar{v}_\infty - v) - \frac{2}{\gamma}\right]q(v) \quad (12b)$$

with the solution

---

[3] The polynomials are normalized according to $\int_0^\infty q(x)\varphi_m(x)\varphi_n(x)dx = \delta_{nm}$

$$q(v) = \frac{\alpha^\alpha}{\Gamma(\alpha)} \frac{v^{\alpha-1}}{\bar{v}_\infty^\alpha} e^{-\alpha \frac{v}{\bar{v}_\infty}}, \qquad (12c)$$

that is the one-parameter Gamma distribution with parameter $\alpha = \gamma \bar{v}_\infty$ and mean value $\bar{v}_\infty$. As a matter of fact, this is the PDF of PV tessellation, attained for KJMA compliant transformations with simultaneous nucleation in the limit $\xi \to \infty$ ($X \to 1$). By inserting in eqn.12c the $\tau$ subscripts previously omitted, the equation reads

$$q_\tau(v) = \frac{\alpha_\tau^{\alpha_\tau}}{\Gamma(\alpha_\tau)} \frac{v^{\alpha_\tau-1}}{\bar{v}_{\tau,\infty}^{\alpha_\tau}} e^{-\alpha_\tau \frac{v}{\bar{v}_{\tau,\infty}}}, \qquad (13)$$

with the $\tau$-dependent parameter of the Gamma distribution $\alpha_\tau = \gamma_\tau \bar{v}_{\tau,\infty}$. In other words, eqn.13 does show that for $t \to \infty$ the population of $\tau$-nuclei are Gamma distributed, with $\tau$-dependent parameter, in accord with the conjecture by Pineda et al [18].

From knowledge of the steady state distribution, it is possible to determine the temporal evolution of the PDF by solving the Sturm-Liouville problem. For the case considered here, again omitting the $\tau$ subscript in $\varphi_{\tau,n}$, $\alpha_\tau$ and $\gamma_\tau$, eqn.11 becomes[4],

$$v \frac{d^2 \varphi_n}{dv^2} + (\alpha - \gamma v) \frac{d\varphi_n}{dv} = -\lambda_n \gamma \varphi_n, \qquad (14a)$$

or, in terms of the new variable $z = \gamma v$,

$$z \frac{d^2 \varphi_n}{dz^2} + (\alpha - z) \frac{d\varphi_n}{dz} = -\lambda_n \varphi_n. \qquad (14b)$$

---

[4] Eqn.6 is rewritten as: $\frac{1}{2} B \varphi'' + A \varphi' + \lambda \varphi = 0$

By considering the solution of eqn.14b in the form of polynomials of degree $n$, the eigenvalues are given by $\lambda_n = n$ (with $n$ positive integer) [23] and the eigenfunctions are the associated Laguerre polynomials $\varphi_n(v) = c_n L_n^{(\alpha-1)}(\gamma v)$, with normalization factor $c_n = \binom{n+\alpha-1}{n}^{-1/2}$, where $\binom{n+\alpha-1}{n} = \frac{\Gamma(\alpha+n)}{n!\,\Gamma(\alpha)}$. With this result eqn.10 can then be written as

$$f(v_0|\xi,v) = q(v) \sum_{n=0}^{\infty} \frac{n!\,\Gamma(\alpha)}{\Gamma(\alpha+n)} e^{-n\xi} L_n^{(\alpha-1)}(\gamma v) L_n^{(\alpha-1)}(\gamma v_0). \qquad (15)$$

We recall that in the KJMA theory the size of the critical nucleus is negligible, which implies considering $v_0 \to 0$ in eqn.15. Since $L_n^{(\alpha-1)}(0) = \frac{\Gamma(\alpha+n)}{n!\,\Gamma(\alpha)}$, the series is equal to $\sum_{n=0}^{\infty} e^{-n\xi} L_n^{(\alpha-1)}(\gamma v) = \frac{1}{(1-\chi)^\alpha} e^{-\gamma v \frac{\chi}{1-\chi}}$, with $\chi = e^{-\xi}$ [24]. By using eqn.13 and eqn.6, the solution for the $\tau$-nuclei is eventually obtained as

$$f_\tau(\xi_\tau, v) = \frac{\alpha_\tau^{\alpha_\tau}}{\Gamma(\alpha_\tau)} \frac{v^{\alpha_\tau-1}}{\bar{v}_\tau(\xi_\tau)^{\alpha_\tau}} e^{-\alpha_\tau \frac{v}{\bar{v}_\tau(\xi_\tau)}}, \qquad (16)$$

or, as a function of $v$ and $t$

$$f_\tau(v,t) = \frac{\alpha_\tau^{\alpha_\tau}}{\Gamma(\alpha_\tau)} \frac{v^{\alpha_\tau-1}}{\bar{v}(t,\tau)^{\alpha_\tau}} e^{-\alpha_\tau \frac{v}{\bar{v}(t,\tau)}}. \qquad (17)$$

According to eqn.17, in terms of the variable $\frac{v}{\bar{v}(t,\tau)}$, the evolution of the PDF is given by the one-parameter Gamma distribution with parameter $\alpha_\tau$. In turn, this parameter stems from the steady-state solution of the FP equation that is the PDF of $\tau$-nuclei at $X \to 1$. For $\tau$-nuclei the value of the parameter $\alpha_\tau$ is therefore kinetically invariant.

As anticipated, the asymptotic Gamma-PDF for each $\tau$-population has originally been hypothesized by Pineda et al [18] and successfully employed for describing computer simulations of KJMA transformations in refs.[18,19]. The present demonstration, based on the theory of stochastic processes, provides a more rigorous support to the Gamma-distribution ansatz. In addition, at finite time the FP solution maintains the same scaling of the asymptotic PDF in the $\frac{v}{\bar{v}}$ time-dependent variables. However, we point out that the finite time solution becomes a better description of the physical system the more the tail of the distribution becomes negligible for $v(t,\tau) > v_e(t,\tau)$. This is due to the presence of a cut off in the PDF linked to the irreversible growth. In terms of the reduced variables defined above, $v_e(t,\tau) = \frac{(t-\tau)^D}{D}$ and from the definition of $\xi_\tau$ we eventually get $v_e(\xi_\tau,\tau) = \left(\frac{\xi_\tau}{b(\tau)}\right)^{\frac{D}{\kappa(\tau)}}$.

The stochastic approach based on the FP equation gives information on the behavior of the fluctuation term, $B_\tau(v)$, that is linear with $v$. In fact, for the Gamma distribution $\alpha_\tau = \frac{(\bar{v}_\tau)^2}{\langle v^2 - (\bar{v}_\tau)^2 \rangle}$ that implies, once inserted in eqn.8b, $\frac{d\Psi_\tau}{d\xi_\tau} = -2\Psi_\tau + 2\frac{\bar{v}_\tau}{\gamma_\tau} = -2\Psi_\tau\left(1 - \frac{\bar{v}_{\tau,\infty}}{\bar{v}_\tau}\right)$, with $\alpha_\tau = \gamma_\tau \bar{v}_{\tau,\infty}$. Use of eqn.5 eventually provides $\frac{d\Psi_\tau}{d\xi_\tau} = 2\Psi_\tau\left(\frac{e^{-\xi_\tau}}{1-e^{-\xi_\tau}}\right)$ with solution $\Psi_\tau = K_\tau(1 - e^{-\xi_\tau})^2$ that is in accord with the definition of $\alpha_\tau$ above, for $K_\tau = (\bar{v}_{\tau,\infty})^2/\alpha_\tau$.

The total PDF is eventually obtained through integration of the $\tau$-nuclei PDF (eqn.17) weighted with the nucleation rate of the actual nuclei:

$$f(v,t) = \frac{\int_0^t I_a(\tau) f_\tau(v,t) d\tau}{\int_0^t I_a(\tau) d\tau}. \qquad (18)$$

In this equation $f(v,t)dv$ is the probability that the size of the nucleus is in the interval between $v$ and $v + dv$ at time $t$ and $I_a(\tau) = (1 - X(\tau))I$ is the actual nucleation rate.

To employ eqns.17 and 18 for describing the PDF of KJMA transformations, the knowledge of the behavior of $\alpha_\tau$ with $\tau$ is needed. As mentioned above $\alpha_\tau$ can be computed at steady state, i.e. at the end of the transformation. To accomplish this, it is profitable to use the analytical approach developed by Pineda et al [18] and by Farjas and Roura [19] which is based on Gilbert's statistical theory. For JM tessellation, that is realized by the growth considered here, the $\alpha_\tau$ parameter is found to decrease with $\tau$ starting from the value $\alpha_0 \cong 12$ at $\tau = 0$ to reach values lower than unity when $X(\tau) \to 1$ [18]. Using these results to estimate $\alpha_\tau$ in the Gamma PDF (eqn.17), the temporal evolution of the total PDF, eqn.18, has been evaluated and compared to computer simulations in Figs.2,3. In the same figures the JM-PDF, attained by solving the FP equation at $b = 0$ [17] (see also the appendix), has also been reported which is in fair agreement with both eqn.18 and the simulations.

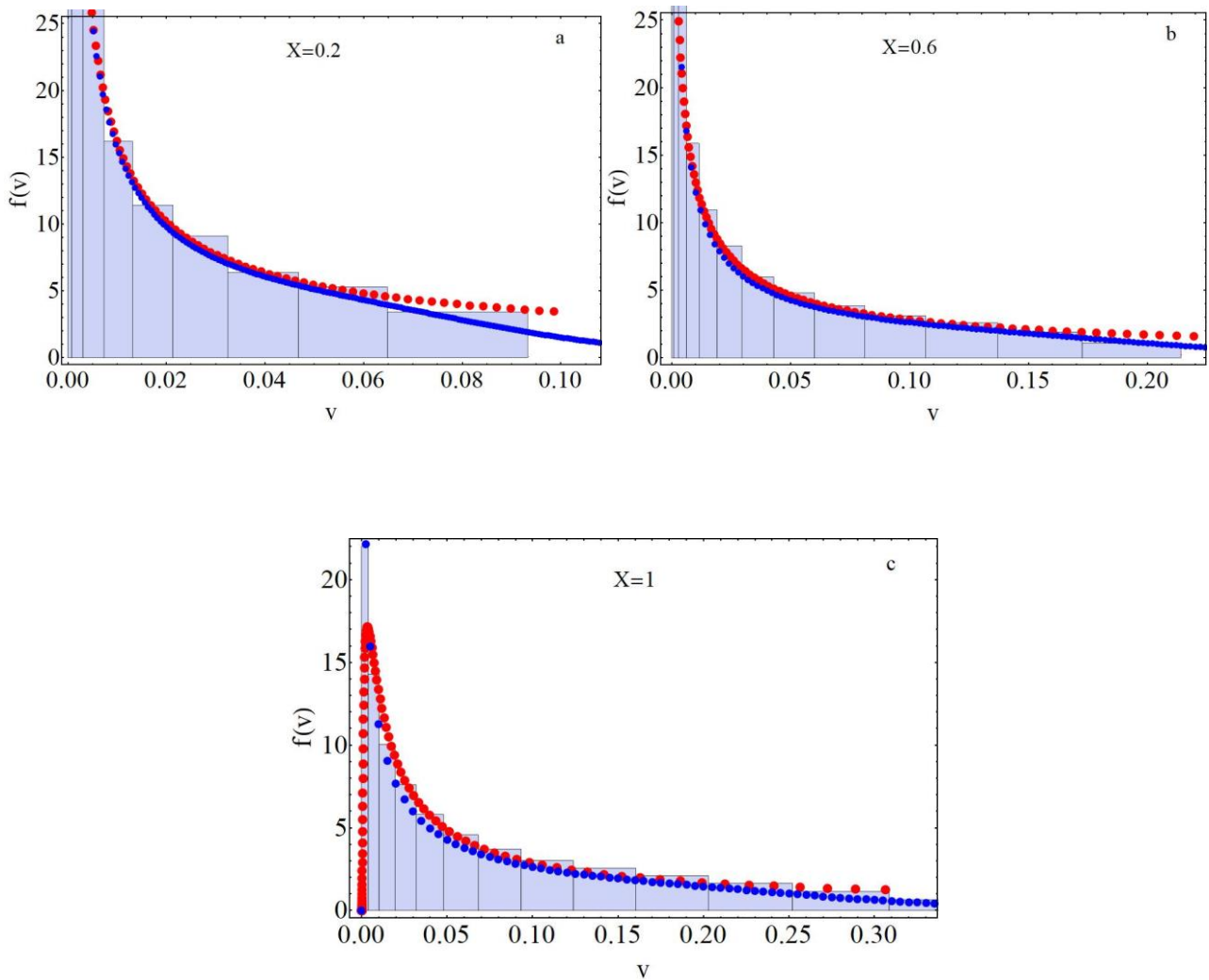

**Fig.2** PDF for KJMA transformation in 3D space with constant nucleation and growth rates. Histograms are the outputs of Monte Carlo simulations for the growth of amorphous Silicon from ref.[6]. The PDF, as a function of volume, has been obtained from the relation $f(v) = F(r)/(4\pi r^2)$. Red and Blue symbols are, respectively, the JM-PDF and the PDF computed by superposition of Gamma-distributed $\tau$-nuclei (eqns.17,18). For the $\tau$-dependent parameter of the Gamma distribution, $\alpha_\tau$, the results of ref.18 were employed.

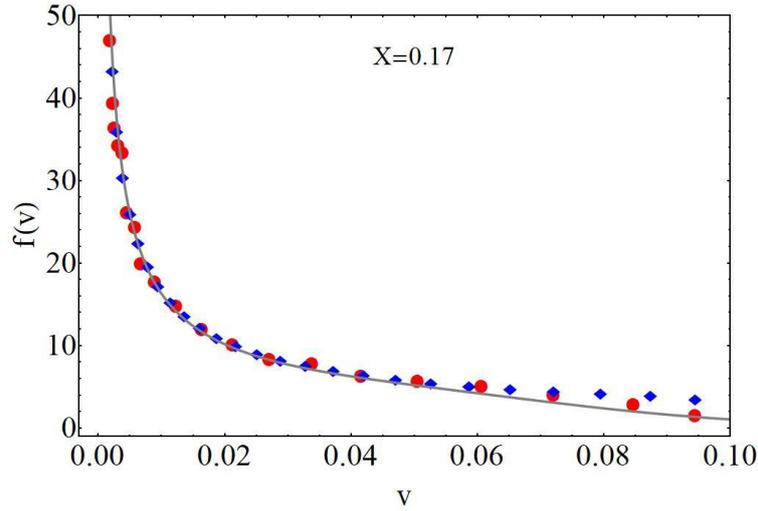

**Fig.3** PDF for KJMA transformation in 3D space with constant nucleation and growth rates. Solid circles are the output of the computer simulation from ref.[25]. The PDF, as function of volume variable, has been obtained from the relation $f(v) = F(r)/(4\pi r^2)$. Full blue diamonds and solid line are, respectively, the JM-PDF and the PDF computed by superposition of Gamma-distributed $\tau$-nuclei (eqns.17,18).

The fact that the JM-PDF provides a good description of the simulations, has already been pointed out in ref.[17]. The present computation provides a rationale to this evidence based on eqn.17 and the behavior of $\alpha_\tau$. This is due to the high value of $\alpha_\tau$, at lower $\tau$, which provides the most important contribution to the phase transformation. In fact, at running time $t$, the contribution to the phase transformation due to the formation of actual nuclei in time interval $0 \leq \tau \leq \tau_c$ (with $\tau_c \leq t$), is equal to $X_t(\tau \leq \tau_c) = \frac{1}{\omega}\int_0^{\tau_c} I_a(z)\bar{x}(t,z)dz$, where reduced time has been used. In terms of the reduced time and volume, defined in section 2.2, the expression reads $X_t(\tau \leq \tau_c) = 12\int_0^{\tau_c} e^{-z^4}\bar{v}(t,z)dz$ [5].

The behavior of the ratio $\eta_t(\tau_c) = \frac{X_t(\tau \leq \tau_c)}{X_t(\tau \leq t)}$ is shown in Fig.4 as a function of $\tau_c$, for linear growth of nucleus radius. In the inset, $X(\tau_c)$ is plotted as a function of $X(t)$ for $\tau_c$ values such that $X_t(\tau \leq \tau_c) = 0.9X_t(\tau \leq t)$. From these curves, we infer that the main contributions to the transformation arise from nucleation events occurring when transformed volume is very low. For instance, nuclei that start growing up to $\tau_c = 0.5$ (i.e. up to $X(\tau_c) = 0.06$), provides 90% of the new phase at $X(t) = 0.98$. This implies that significant contributions to the total PDF arise from

---

[5] In terms of reduced quantities $X(t) = D(D+1)\int_0^t e^{-z^4}\bar{v}(t,z)dz$, similarly for $X_t(\tau \leq \tau_c)$.

Gamma distributions with higher $\alpha$ values where the PDF mimics a Dirac delta function. This justifies the good agreement between the JM-PDF and the computer simulations.

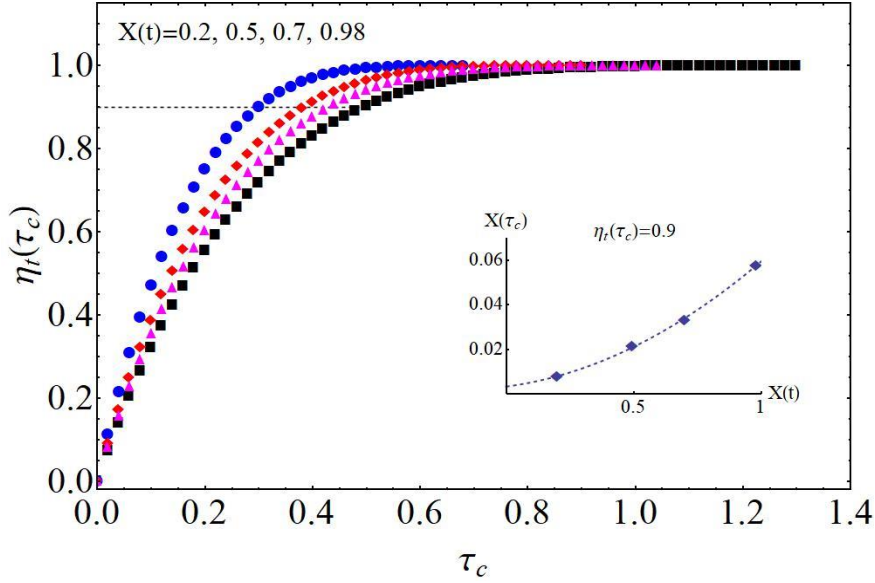

**Fig.4** Contribution of the $\tau$-populations, with $\tau \in [0, \tau_c]$, to the fraction of transformed volume at running time $t$: $\eta_t(\tau_c) = \frac{X_t(\tau \leq \tau_c)}{X_t(\tau \leq t)}$ where $X_t(\tau \leq \tau_c)$ is the fraction of transformed volume, at time t, due to all populations with $0 \leq \tau \leq \tau_c$. Circle $X(t) = 0.2$; diamond $X(t) = 0.5$; triangle $X(t) = 0.7$; square $X(t) = 0.98$. In the inset, the value $X(\tau_c)$ is reported as a function of $X(t)$ for $\tau_c$ values that provides a contribution equal to 90% of $X(t)$.

For the sake of completeness, we point out that, setting $B(\xi_\tau, v) = B(\xi_\tau)$ in eqn.7b, we would obtain a Gaussian-like solution with variance given by the solution of eqn.8b [10,17]. However, at lower values of the transformed volume, the Gaussian-like distribution is an approximation since the inequality $\sqrt{\Psi_\tau} \ll \bar{v}_\tau$ is not satisfied and $f(0,t)$ is non zero. According to this argument, in the series expansion of $B(\xi_\tau, v)$, the first term $B(\xi_\tau, 0)$, is negligible, at least at early times.

Before concluding this section, we briefly comment on the non-isothermal behavior of the kinetics, with constant heating rate, $\phi$, where the temperature is proportional to time: $T(t) = \phi t$. Even for this case, in terms of a new time dependent variable, the kinetics of the mean volume can be expressed by an equation like eqn.6. For non-isothermal transformations the temperature dependent terms bring an extra time dependence in the equations for the transformed volume and mean volume of nuclei. Changing the integration variables from time to temperature and introducing the heating rate $\phi = \frac{dT}{dt}$ (a constant), the equations for the mean volume and the extended volume become:

$$\bar{v}(T_t, T_\tau) = \frac{D g_D G_0^D}{\phi^D} \int_{T_\tau}^{T_t} dz \ e^{-[X_e(z) - X_e(T_\tau)]} e^{-E_g/kz} \left[ \int_{T_\tau}^{z} dT'' e^{-E_g/kT''} \right]^{D-1}, \quad (19)$$

$$X_e(T_t) = \frac{I_0 g_D G_0^D}{\phi^{D+1}} \int_{T_0}^{T_t} dT' \ e^{-E_n/kT'} \left[ \int_{T'}^{T_t} dT'' e^{-E_g/kT''} \right]^{D}, \quad (20)$$

where $E_n$ and $E_g$ are the activation energies for nucleation and growth, respectively. The nucleation and growth rates are $I(T) = I_0 \ e^{-E_n/kT}$ and $\frac{dr}{dt} = G_0 e^{-E_g/kT}$, respectively. The behavior of the mean volume, eqn.19, has been displayed in Fig.5 as a function of actual temperature, $T_t$ and for different values of $T_\tau$, that is the temperature at which the $\tau$-population starts growing. In the figure, solid lines are the best fit of a stretched exponential function, like eqn.3, to the mean volume. The good agreement between the curves and the data set, motivates the use of the FP equation also for the isochronal transition employing the same approach above discussed for the isothermal process. It follows that the PDF is given by the superposition of Gamma functions with parameters computed using the statistical approach of ref.[20]. Notably, for non-isothermal transformations this parameter has been estimated by Farjas and Roura [19] and follows a similar behavior as the isothermal case: at the beginning of the transition the parameter is $\alpha_0 \cong 14$ and decreases with the increase of $\tau$. Consequently, like the isothermal process, the PDF for the isochronal transition should match the JM-PDF given by eqn.3b and eqns.19, 20. In fact, in ref.[17] the agreement between the two PDF's has been successfully checked for the crystallization of amorphous Silicon.

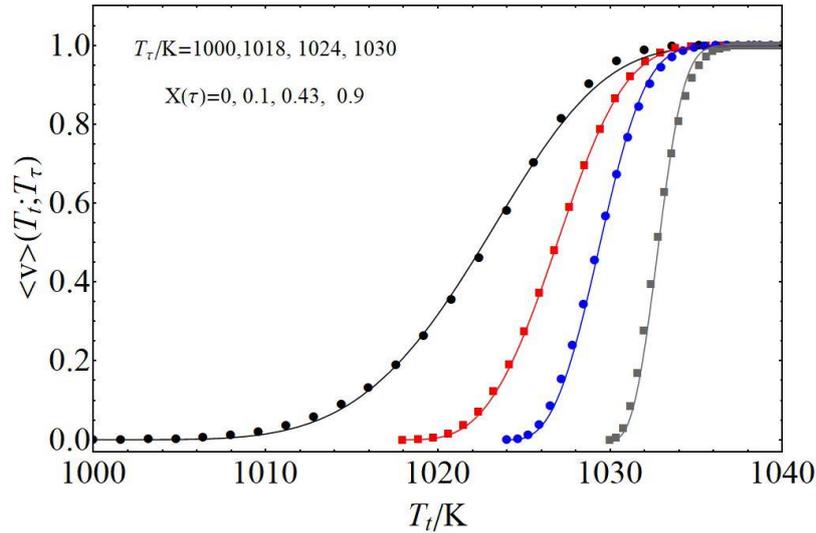

**Fig.5** Mean particle volume for non-isothermal JKMA transformation (eqns.19,20). Computations of mean volume for isochronal crystallization of amorphous Silicon are displayed as symbols (parameter values from ref.19). Solid lines are the best fit of the stretched exponential function to the numerical output. Mean volumes have been normalized to the asymptotic values.

**Conclusions**

The Fokker-Planck equation for the particle size PDF in the classical KJMA transformations has been established based on the rate equation for the mean volume of particles, $\bar{v}(t,\tau)$. A suitable choice of the time scale allows one to find the expression for the "drift term" in the FP equation $A(\xi,v)$. For a linear dependence of the "diffusion term" with particle size, the solution of the FP equation for each $\tau$-population is the one parameter Gamma PDF with mean value $\bar{v}(t,\tau)$. The parameter of the Gamma distribution, $\alpha_\tau$, is shown to be invariant during the transformation and, consequently, can be estimated from the steady state solution of the FP equation, that is at $X=1$. The present approach, based on the FP equation, offers a demonstration of the conjecture proposed by Pineda et al [18] on the PDF of the $\tau$-population of particles. The final PDF is eventually computed by the convolution product of the $\tau$-PDF with the actual nucleation rate, where the parameters of the Gamma distributions have been attained employing the method of ref.[18]. The theoretical PDF is in good agreement with Monte Carlo simulations of the particle size distribution functions for constant rates of both nucleation and growth. Moreover, in the volume domain the PDF is shown to be in accord with the JM-PDF. The FP solution provides a rationale for this behavior that is due to the large $\alpha_\tau$ values at the beginning of the transformation where the contribution of the $\tau$-populations to the transformation is significant. It is also shown that similar argument holds for non-isothermal transformations given a constant rate of heating.

**Appendix**

*A- Johnson-Mehl particle size distribution function*

Let $g(v,t)$ be the number density of nuclei with size between $v$ and $v+dv$ at $t$. Conservation of number density of nuclei reads

$$\int_0^\infty g(v,t)dv = \int_0^t I_a(t')dt' . \qquad (A1)$$

In terms of $f(\tau, 0|t, v)$ this function is obtained by summing over the birth time of populations of nuclei:

$$g(v,t) = \int_0^t I_a(\tau) f(\tau, 0|t, v) d\tau. \qquad (A2)$$

By neglecting fluctuations around the mean size, one gets $f(\tau, 0|t, v) = \delta(v - \bar{v}(t,\tau))$. Eqn. A2 provides

$$g(v,t) = \int_0^t I_a(\tau) \delta(v - \bar{v}(t,\tau)) d\tau, \qquad (A3)$$

Because of $\delta(v - \bar{v}(t,\tau)) = \delta(\tau - \tau(v,t))/\left|\frac{\partial \bar{v}(t,\tau)}{\partial \tau}\right|$, with $\tau(v,t)$ the root of the equation $v = \bar{v}(t,\tau)$, eqn. A3 becomes

$$g(v,t) = \frac{I_a[\tau(v,t)]}{\left|\frac{\partial \bar{v}(t,\tau)}{\partial \tau}\right|_{\tau(v,t)}} . \qquad (A4)$$

Eqn. A1 provides the normalization factor $N = \int_0^t I_a(t')dt'$. In terms of the reduced variables, defined in section 2.1, the actual nucleation rate is $I_a(\tau) = (1 - X(\tau))I = I e^{-\tau^{D+1}}$ and the normalized PDF is given by

$$f_{JM}(v,t) = \frac{e^{-\tau(v,t)^{D+1}}}{\int_0^t e^{-\xi^{D+1}}d\xi} \left(\left|\frac{\partial \bar{v}(t,\tau)}{\partial \tau}\right|_{\tau(v,t)}\right)^{-1}, \qquad (A5)$$

that is eqn. 3b. Finally, by using the time evolution of the mean size, eqn. 3a, we obtain

$$f_{JM}(v,t) = N^{-1} \frac{e^{-(\tau)^{D+1}}}{e^{(\tau)^{D+1}}(D-1)\int_\tau^t e^{-(z)^{D+1}}(z-\tau)^{D-2}dz - (D+1)\tau^D v}, \qquad (A6)$$

where $\tau \equiv \tau(v,t)$ is given by inverting eqn. 3a.

*B-derivation of eqns.8a and 8b*

We consider eqn.7b,

$$\frac{\partial f(\xi,v)}{\partial \xi} = -\frac{\partial [A(\xi,v)f(\xi,v)]}{\partial v} + \frac{1}{2}\frac{\partial^2 [B(\xi,v)f(\xi,v)]}{\partial v^2}, \quad (B1)$$

where the subscript $\tau$ has been omitted. The rate equation for the mean volume is obtained by using eqn.B1 in the expression

$$\frac{d\langle v \rangle}{d\xi} = \frac{\partial}{\partial \xi} \int_0^\infty vf(\xi,v)dv = \int_0^\infty v\frac{\partial f}{\partial \xi}dv,$$

that is

$$\frac{d\langle v \rangle}{d\xi} = -\int_0^\infty v\frac{\partial [A(\xi,v)f(\xi,v)]}{\partial v}dv + \frac{1}{2}\int_0^\infty v\frac{\partial^2 [B(\xi,v)f(\xi,v)]}{\partial v^2}dv. \quad (B2)$$

An integration by parts leads to

$$\frac{d\langle v \rangle}{d\xi} = \int_0^\infty A(\xi,v)f(\xi,v)dv + \frac{1}{2}\left[v\left(\frac{\partial (Bf)}{\partial v} - 2Af\right) - Bf\right]_0^\infty = \langle A \rangle. \quad (B3)$$

Similar computation pathway is employed for the mean of the square of the volume. A double integration by parts provides

$$\frac{d\langle v^2 \rangle}{d\xi} = \left[-v^2\left(Af - \frac{1}{2}\frac{\partial (Bf)}{\partial v}\right) - vBf\right]_0^\infty$$
$$+ 2\int_0^\infty vA(\xi,v)f(\xi,v)dv + \int_0^\infty B(\xi,v)f(\xi,v)dv = 2\langle vA \rangle + \langle B \rangle. \quad (B4)$$

In eqns.B3, B4 the terms in the square brackets vanish at the extremes of integration. Furthermore, eqn.8b is eventually obtained by using eqns.B3, B4 in the rate equation for the variance: $\frac{d\langle v^2 - \langle v \rangle^2 \rangle}{d\xi} = \frac{d\langle v^2 \rangle}{d\xi} - 2\langle v \rangle \frac{d\langle v \rangle}{d\xi}$.

# References

1) A.N. Kolmogorov, Izv. Akad. Nauk SSSR Ser. Mat. **3** (1937) 355

2) M. Avrami, J. Chem. Phys. **7** , (1939),1103 ; ibid. J. Chem. Phys. **8** (1940) 212  ibid. J. Chem. Phys. **9** , (1941), 177

3) W.A. Johnson, R.F. Mehl, Trans. Trans. Am. Inst. Min. (Metall.) Eng. **135** (1939) 416; K. Barmak, Met. and Mater. Trans. **41**A (2010) 2711;

4) N. V. Alekseechkin, J. Non Cryst. Solids **357** (2011) 3159

5) J.S. Blazquez, F.J. Romero, C.F. Conde, A. Conde, Phys. Status Solidi (B), **256** (6) 2100524

6) E. Pineda, D. Crespo, Phys. Rev. B **60** (1999) 3104

7) E. Pineda, T. Pradell, D. Crespo, J. Non Cryst. Solids **287** (2001) 88

8) J. Farjas, P. Roura, Phys. Rev. B **75** (2007) 184112

9) E.Pineda, D. Crespo, J. of Statist. Mechanics: theory and Experiment 2007 (2007) P06007

10) V. G. Dubrovskii, The J. of Chem. Phys. **131** (2009) 164514

11) V. G. Dubrovskii, M. V. Nazarenko, The J. of Chem. Phys. **132** (2010) 164514

12) D. Hömberg, F.V. Patacchini, K. Sakamoto, J. Zimmer,  IMA Journal of Applied Mathematics (Institute of Mathematics and Its Applications), **82**(4) (2017) 763

13) Z. Néda, F. Jarai-Szabo, S. Boda, Phys. Rev.E **96** (2017) 042145

14) M. Tanemura, Forma **18** (2003) 221

15) T. Kiang, Z. Astrophys. **64** (1966) 433

16) J.-S. Ferenc, Z. Néda, Physica A **385** (2007) 518

17) M. Tomellini, J. Cryst. Growth, **584** (2022) 126579

18) E. Pineda, P. Bruna, D. Crespo, Phys. Rev. E **70** (2004) 066119

19) J. Farjas, P. Roura, Phys. Rev. B **78** (2008) 144101

20) E. Gilbert, Ann. Math. Stat. **33** (1962) 958.

21) N.G. Van Kampen, *Stochastic Processes in Physics and Chemistry*, North Holland Publishing Company, 1981 , Amsterdam, New York, Oxford.

22) E. Wong, J.B. Thomas, J. Soc. Indust. Appl. Math. **10** (1962) 507

23) The Sturm-Liouville problem is, $\frac{1}{2}B\varphi''(x) + A\varphi'(x) + \lambda\varphi(x) = 0$. For the case under study $A$ and $B$ are in the form $A = ax + b$, $B = cx$ (with constant $a, b,$ and $c$) The polynomial solution of degree $n$, reads $\varphi_n(x) = \sum_{k=0}^{k=n} d_k^{(n)} x^k$, that is further inserted in the differential equation for determining the coefficients $d_k^{(n)}$. This is done by setting equal to zero each coefficient of the polynomial resulting in the first member of the equation. The coefficient of the term $x^n$ is equal to $[n\,a + \lambda]d_k^{(n)}$; it is zero for $\lambda = -an$, namely for $\lambda = n$ since $a = -1$ in our case (section 2.2).

24) I.S. Gradshteyn and I.M. Ryzhik, *Table of Integrals, Series, and Products* Seventh Edition, Academic Press (2007)

25) M.R. Riedel, S. Karato, Geophys. J. Int. **125** (1986) 397